\documentclass[fdp,a4paper,fleqn%
]{w-art}
\usepackage{times,cite,w-thm}
\theoremstyle{plain}

\theoremstyle{definition}

\usepackage[]{graphicx}
\begin{document}
\DOIsuffix{theDOIsuffix}
\Volume{55}
\Month{01}
\Year{2007}
\pagespan{1}{}
\Receiveddate{}
\Reviseddate{}
\Accepteddate{}
\Dateposted{}



\title{Vector Supersymmetry from $\rm OSp(3,2|2)$: Casimir operators}


\author[Casalbuoni]{Roberto Casalbuoni\inst{1,}\footnote{Roberto Casalbuoni\hspace{.36 cm} E-mail:~\textsf{casalbuoni@fi.infn.it},
            \quad \quad \quad\quad \quad \quad\quad \quad \,\,\hspace{.7 mm}Phone: +39\,55\,4572318\,
            \,\hspace{.2 mm} Fax: +39\,55\,4572121\,}}
\address[\inst{1}]{Dipartimento di Fisica, Universit\`a di Firenze, INFN-Firenze and Istituto Galileo Galilei, Via Sansone 1,
50019 Sesto Fiorentino (FI), Italy}
\author[Elmetti]{Federico Elmetti\inst{2,}%
  \footnote{Federico Elmetti\quad \,\,\,\,\,\,\,\,E-mail:~\textsf{federico.elmetti@fys.kuleuven.be},
            \quad \quad \,\,\,\hspace{.3 mm} Phone: +32\,16\,327231\,
            \quad Fax: +32\,16\,327986}}
\address[\inst{2}]{Instituut voor Theoretische Fysica, Katholieke Universiteit Leuven, Celestijnenlaan 200D, B-3001 Leuven, Belgium}
\author[Gomis]{Joaquim Gomis\inst{3,}\footnote{Joaquim Gomis\quad \hspace{.27 cm} E-mail:~\textsf{gomis@ecm.ub.es},
            \quad \quad \quad \quad \quad \quad \quad \quad \quad \,\hspace{.39 mm}Phone: +34\,934\,021177\,
            \,\hspace{.28 mm} Fax: +34\,934\,021198}}
\address[\inst{3}]{Departament d'Estructura i Constituents de la Mat{\`e}ria and Institut de Ci{\`e}ncies del Cosmos, Universitat de Barcelona,
Diagonal 647, 08028 Barcelona, Spain}
\author[Kamimura]{Kiyoshi Kamimura \inst{4,}\footnote{Kiyoshi Kamimura\quad \,\,\,\,\,E-mail:~\textsf{kamimura@ph.sci.toho-u.ac.jp},
            \quad \quad \quad \,\,\,\,\,\,\hspace{.23 mm}Phone: +81\,47\,472\,6876\,
            \hspace{.25 mm} Fax: +81\,47\,475\,1855}}
\address[\inst{4}]{Department of Physics, Toho University, Funabashi, 274-8510 Japan}
\author[Tamassia]{Laura Tamassia \inst{2,}\footnote{Laura Tamassia \hspace{0.77 cm} E-mail:~\textsf{laura.tamassia@fys.kuleuven.be},
            \quad \quad \quad \,\hspace{.1 mm}Phone: +32\,16\,327249\,
            \quad Fax: +32\,16\,327986}}
\author[Van Proeyen]{Antoine Van Proeyen \inst{2,}\footnote{Antoine Van Proeyen\quad E-mail:~\textsf{antoine.vanproeyen@fys.kuleuven.be},
            \quad Phone: +32\,16\,327240\,
            \quad Fax: +32\,16\,327986}}

\begin{abstract}

In this paper we briefly review the main results obtained in \cite{ourpaper}, where some algebraic properties of the `vector supersymmetry' (VSUSY) algebra have been studied. VSUSY is a graded extension of the Poincar\'e algebra in 4 dimensions with two central charges. We derive all independent Casimir operators of VSUSY and we find two distinct spin--related operators in the case of nonvanishing central charges. One is the analogue of superspin for VSUSY and the other is a new spin, called C--spin, whose value is fixed to $1/2$.
We also show that the VSUSY algebra and its Casimir operators can be derived by an In{\"o}n{\"u}-Wigner contraction from $\rm OSp(3,2\,|2)$. This paper is based on the talk given in Varna, Bulgaria, during the 4-th EU RTN Workshop 2008.

\end{abstract}

\maketitle                   






\section{Introduction}

In \cite{ourpaper} we initiated a systematic study of the algebraic properties of vector supersymmetry (VSUSY),
a graded extension of the Poincar{\'e} algebra in four dimensions. In contradistinction to ordinary supersymmetry, the fermionic generators of VSUSY are not spinors. They are an odd Lorentz vector $G_{\mu}$  and an odd Lorentz scalar $G_{5}$. Moreover, two central charges, $Z$ and $\tilde Z$, are allowed. The name vector supersymmetry comes from the fact that the anticommutator between  $G_{\mu}$  and a $G_{5}$ gives the four-momentum vector $P_{\mu}$. This algebra was first introduced in \cite{Barducci:1976qu} in 1976 with the purpose of obtaining a pseudoclassical description of the Dirac equation.

Since the odd generators of VSUSY are a vector and a scalar, they do not satisfy the spin-statistics rule.
This particular feature of VSUSY is not necessarily an obstacle to the construction of physical models with underlying VSUSY. For example, a quantum-mechanical model for the spinning particle with underlying VSUSY was first constructed in \cite{Casalbuoni:2008iy}.

VSUSY is not only worth exploring as a new kind of supersymmetry, but it has also an interesting connection with topological field theories. In fact, an Euclidean version of VSUSY appears as a subalgebra of the symmetry algebra underlying topological $\mathcal{N}=2$ Yang-Mills theories, in the special case when the two VSUSY central charges $Z$ and $\tilde Z$ are identified \cite{Witten:1988ze,Alvarez:1994ii,Kato:2005fj}.

In order to understand the physical content of VSUSY, we need to classify its irreducible representations. A first step in this direction was taken in \cite{ourpaper}, where all the Casimir operators of the algebra were constructed. VSUSY has four even Casimirs, $P ^2$, $Z$, $\tilde{Z}$ and $\hat{W}^2$. The latter is the square of $\hat{W}_{\mu}$, the analogue of the superspin vector of ordinary supersymmetry \cite{Buchbinder:1998qv}. $\hat{W}_{\mu}$ is constructed as a combination of the Pauli-Lubanski vector $W_{\mu}$ and another vector, called $W_{C\,\mu}$. Like $\hat{W}_{\mu}$, $W_{C\,\mu}$ satisfies, in the rest frame, the $\rm SU(2)$ algebra and therefore defines a new spin, called C-spin. The operator $W_{C}^{2}$ is a Casimir but it is not independent since it is proportional to $P^{2}$. As a result, the value of the C-spin is fixed to $1/2$. This feature is a peculiarity of VSUSY and holds for the generic case with nonvanishing central charges.

The fact that the odd generators of VSUSY are not spinors but a vector and a scalar allows for the existence of odd Casimir operators. In fact, we have constructed an odd operator that commutes with all generators of VSUSY only when a particular BPS-like relation between the 4-momentum and the central charges is satisfied.

Finally, we briefly review the connection between VSUSY and the simple orthosymplectic superalgebra $\rm OSp(3,2\,|2)$ by showing how to derive both VSUSY and its Casimirs by contraction from $\rm OSp(3,2\,|2)$.

The paper is organized as follows: In Section 2 we introduce the VSUSY algebra and we discuss its Casimir operators. In Section 3 we introduce the $\rm OSp(3,2\,|2)$ algebra and we show how to derive VSUSY by a contraction procedure. In the same section we also discuss the contraction of the Casimir operators and specially how to derive the analogue of superspin for VSUSY. In Section 4, we summarize our results and present our plans for future work.

\section{VSUSY algebra and its Casimir operators}

The vector supersymmetry (VSUSY) algebra is a graded extension of the Poincar\'e algebra. In addition to the Poincar\'e generators $M_{\mu\nu}$ and $P_{\mu}$, two fermionic generators $G_{\mu}$ and $G_{5}$ and two bosonic central charges $Z$ and $\tilde Z$ are present. $G_{\mu}$ and $G_{5}$ behave respectively as a four-vector and a scalar under the Lorentz group. The VSUSY algebra reads:
\begin{equation}
[M_{\mu\nu},M_{\rho\sigma}]_{-}=\eta_{\nu\rho}M_{\mu\sigma}+
\eta_{\mu\sigma}M_{\nu\rho}-\eta_{\nu\sigma}M_{\mu\rho}-
\eta_{\mu\rho}M_{\nu\sigma}, \nonumber
\label{lorentz}
\end{equation}
\begin{equation}
[M_{\mu\nu},P_\rho]_{-}=-\eta_{\mu\rho}P_\nu+\eta_{\nu\rho}P_\mu\,,\nonumber
\label{poincare}
\end{equation}
\begin{equation}
[M_{\mu\nu},G_\rho]_{-}=-\eta_{\mu\rho}G_\nu+\eta_{\nu\rho}G_\mu\,,\nonumber
\end{equation}
\begin{equation}
[G_\mu,G_\nu ]_+=\eta_{\mu\nu}Z\,,\quad [G_5,G_5 ]_+=\tilde Z
\,,\quad [G_\mu,G_5]_+=-P_\mu\,.
\label{vsusyalgebra}
\end{equation}
A systematic procedure to find all the even Casimirs of VSUSY is to start from the most general form of an even scalar operator:
\begin{eqnarray}
\mathcal{C}=C+C^{\mu\nu}G_{\mu}G_{\nu}+C^{\mu 5} G_{\mu}G_{5}
+C^{*}\epsilon^{\mu\nu\rho\sigma}G_{\mu}G_{\nu}G_{\rho}G_{\sigma}+
C^{*}_{\mu}\epsilon^{\mu\nu\rho\sigma}G_{\nu}G_{\rho}G_{\sigma}G_{5},
\label{CCgeneral}
\end{eqnarray}
where the coefficients $C$'s are functions of the bosonic generators $(P,M,Z,\tilde Z)$ and $C^{\mu\nu}$ is antisymmetric.\\
By imposing that $[\mathcal{C},G_{5}]_{-}=0$, the form of $\mathcal{C}$ gets further constrained:
\begin{eqnarray}
\mathcal{C}&=&
C+C^{\mu\nu}\tilde G_{\mu}\tilde G_{\nu}+C^*\epsilon^{\mu\nu\rho\sigma}\tilde G_{\mu}\tilde
G_{\nu} \tilde G_{\rho} \tilde G_{\sigma}\, , \label{CCCgen}
\end{eqnarray}
where  we define $\tilde G_{\mu}\equiv G_{\mu}+\frac{1}{\tilde Z}P_{\mu} G_{5}$.
The invariance of $\mathcal{C}$ with respect to the Poincar{\'e} subgroup implies
that the $C$'s transform as Lorentz covariant tensors and that they are
functions of $P_{\mu}$, $W_{\mu}$, $Z$ and $\tilde Z$, so that
\begin{eqnarray}
&&C=C(P^{2},W^{2},Z,\tilde Z),\qquad C^{*}=C^{*}(P^{2},W^{2},Z,\tilde Z)\,,\nonumber\\
&&C^{\mu\nu}=C'(P^{2},W^{2},Z,\tilde Z)\epsilon^{\mu\nu\rho\sigma}P_{\rho} W_{\sigma}+
C''(P^{2},W^{2},Z,\tilde Z)P^{[\mu} W^{\nu]}\,, \label{CCC}
\end{eqnarray}
where $W^{\mu}$ is the Pauli-Lubanski vector
\begin{equation}
W^{\mu}= \frac{1}{2}\epsilon^{\mu\nu\rho\sigma}P_{\nu} M_{\rho\sigma}.
\end{equation}
As in the case of ordinary supersymmetry, the square of the Pauli-Lubanski vector is not a Casimir. This simply means that particles with different Lorentz spin will appear in the same supermultiplet.\\
It is straightforward to see that we have three Casimir operators which are independent of the $\tilde G$'s:
\begin{equation}
P^{2}\,,\quad Z\,\quad \mbox{and}\quad  \tilde Z\,.
\end{equation}
After some algebraic manipulations, one can show that an even Casimir containing two $\tilde G$'s must have the following form:
\begin{eqnarray}
\mathcal{C}_{(2)}=ZW^2+\epsilon^{\mu\nu\rho\sigma}P_{\rho} W_{\sigma} \tilde G_{\mu} \tilde G_{\nu}\,=ZW^2+\epsilon^{\mu\nu\rho\sigma}P_{\rho} W_{\sigma} G_{\mu} G_{\nu}\,.
\label{C2G4}\end{eqnarray}
If one introduces a VSUSY generalization of the Pauli-Lubanski vector\footnote{In the ordinary supersymmetry case, the correct generalization of the Pauli-Lubanski vector is $\tilde W^{\mu}\equiv W^{\mu}-\frac{1}{2}[Q,\bar{Q}]\sigma^{\mu}$
and the relative superspin Casimir is $\mathcal{C}=\left(\tilde W^{[\mu}P^{\nu]}\right)^{2}$.} as follows:
\begin{eqnarray}
\hat W^{\mu}& \equiv & ZW^{\mu}-\frac{1}{2}
\epsilon^{\mu\nu\rho\sigma}P_{\nu} G_{\rho} G_{\sigma},
\label{superspinIIA}
\end{eqnarray}
one can show that its square is given by
\begin{eqnarray}
\hat W^{2}=Z\mathcal{C}_{(2)} + \frac{3}{4}P^{2} Z^{2}\,,
\label{superspincasimir}
\end{eqnarray}
which is therefore a Casimir. One can prove that the second term on the RHS of (\ref{superspinIIA})
\begin{equation}
W_{C}^{\mu}\equiv \frac{1}{2}
\epsilon^{\mu\nu\rho\sigma}P_{\nu} G_{\rho} G_{\sigma}
\end{equation}
also leads to a Casimir, since its square is
\begin{equation}
W_{C}^{2}=\frac{3}{4}\,Z^2 P^2\;.
\label{C-spin}
\end{equation}
This is clearly not an independent Casimir.\\
The three vectors $W_{*}^{\mu}=\frac{\hat W^{\mu}}{Z}$, $W^{\mu}$, $\frac{W_{C}^{\mu}}{Z}$
satisfy the commutation relations
\begin{eqnarray}
\left[W_{*}^{\mu},W_{*}^{\nu}\right]_{-}=\epsilon^{\mu\nu\rho\sigma}P_{\rho}
W_{*\sigma}\,.\label{spinalgA}
\end{eqnarray}
Therefore, in the rest frame of the massive states where $P^{2}=-m^2$, they satisfy the rotation algebra
\begin{eqnarray}
 \left[\frac{W_{*}^{i}}{m},\frac{W_{*}^{j}}{m} \right]_{-}=\epsilon^{ijk}\,\frac{W_{*k}}{m}
\end{eqnarray}
and define three different spins. The superspin $Y$ labels the eigenvalues $- m^2Z^2Y(Y+1)$ of the Casimir ${\hat W}^2$.
The spin associated to $W_C^2$ (C-spin) is fixed to $1/2$, as one can see from (\ref{C-spin}).
Finally, we denote the usual Lorentz spin by $s$.

A few comments are in order. First, from (\ref{superspincasimir}) it is clear that the superspin Casimir ${\hat W}^2$ collapses to $W_{C}^{2}$ in the case of vanishing central charges. Second, it can be shown there are no even Casimir operators containing more than two $\tilde G$'s.\\
We summarize our results and compare to ordinary supersymmetry in the following tables.\vskip 3mm
\begin{center}
\hspace{-.5mm}\begin{tabular}{|c|c|c|c|}
 \hline
 & (Super)spin operator & Casimir & $|\rm Eigenvalue|$\\
 \hline
 & & &\\
 Poincar\'e &$ W^{\mu}= \frac{1}{2}\epsilon^{\mu\nu\rho\sigma}P_{\nu} M_{\rho\sigma} $ & $\mathcal{C}=W^{2} $ & $m^2\;s(s+1)$ \\
 & & &\\
 Super Poincar\'e & $ \tilde W^{\mu}\equiv W^{\mu}-\frac{1}{2}[Q,\bar{Q}]\sigma^{\mu} $ & $\mathcal{C}=\left(\tilde W^{[\mu}P^{\nu]}\right)^{2}$ & $m^{4} S(S+1)$\\
 & & &\\
 VSUSY & $\hat W^{\mu}=ZW^{\mu}-W_{C}^{\mu}$ & $\mathcal{C}=\hat W^{2}$ & $m^{2}Y(Y+1)$\\
 & & &\\
 \hline
\end{tabular}
\vskip 3mm
\textbf {Table 1}: (Super)spins for Poincar\'e, Super Poincar\'e and VSUSY
\vskip 6mm

\hspace{-.5mm}\begin{tabular}{|c|c|c|c|}
 \hline
 & C-spin operator & Casimir & $|$ C-spin $|$ \\
 \hline
 & & &\\
 Poincar\'e & $  - $ & $ - $ & $ - $ \\
 & & &\\
 Super Poincar\'e & $ \tilde W^{\mu}\equiv \frac{1}{8}\bar Q\gamma^{\nu}\gamma_{5}Q\left(\delta_{\nu}^{\mu}-\frac{P_{\nu}P^{\mu}}{P^{2}}\right) $ & $ \rm \tilde W^{2}\,is\,not\, a\, Casimir $ & \rm State-dependent \\
 & & &\\
 VSUSY & $ W_{C}^{\mu}\equiv \frac{1}{2}
\epsilon^{\mu\nu\rho\sigma}P_{\nu} G_{\rho} G_{\sigma} $ & $\mathcal{C}=W_{C}^{2}=\frac{3}{4}\,Z^2 P^2$ & $ 1/2 $\\
 & & &\\
 \hline
\end{tabular}
\vskip 3mm
\textbf {Table 2}: C-spin for Poincar\'e, Super Poincar\'e and VSUSY
\end{center}
Let us now turn to the construction, if possible, of odd Casimir operators. For ordinary supersymmetry, there can be no odd Casimirs, as the fermions are spinors and hence do not commute with the Lorentz generators. However, for VSUSY this argument does not hold. In fact, one can show that the following odd operator
\begin{equation}
Q\equiv G^{\mu}P_{\mu} +G_{5}Z
\end{equation}
commutes with all generators of the algebra when the following condition between the 4-momentum and the central charges
is satisfied:
\begin{equation}
Z\tilde Z = P^{2}.
\label{bps}
\end{equation}
Finally, it can be shown that there are no odd Casimir operators which are cubic or of higher order in the odd generators.

\section{VSUSY algebra and Casimir operators from $\rm OSp(3,2\,|2)$}

In this section we discuss how the VSUSY algebra arises as a subalgebra of an In{\"o}n{\"u}-Wigner contraction of the simple orthosymplectic algebra $\rm OSp(3,2\,|2)$. $\rm OSp$ algebras are natural candidates for an embedding of VSUSY. In fact, for generic $\rm OSp(M|N)$ algebras, the fermions are vectors of $\rm SO(N)$ and $\rm Sp(M)$ and we precisely need our VSUSY fermions to appear as vectors (or scalars) of the Lorentz group. We will use the embedding in $\rm OSp(3,2\,|2)$, whose bosonic part is $\rm SO(3,2)\times \rm Sp(2)$. The latter factor will host the central charges $Z$ and $\tilde Z$ we want to include.

The bosonic generators of  $\rm OSp(3,2\,|2)$ are $M_{\mu\nu}$, $P_{\mu}$, $Z$, $\tilde{Z}$ and $Z'$, while the fermionic generators are $G_{\mu}$,  $S_{\mu}$,  $G_{5}$ and $S_{5}$. The subset of generators $(M_{\mu\nu},P_{\mu}, Z, \tilde{Z}, G_{\mu}, G_{5})$ is the one of interest since it will give rise, after a proper contraction, to the VSUSY algebra. The commutation relations for this subsector are
\begin{eqnarray}
&&\left[M_{\mu\nu},M_{\rho\sigma}\right]_{-}=\eta_{\nu\rho}M_{\mu\sigma} + \eta_{\mu\sigma}M_{\nu\rho} - \eta_{\mu\rho}M_{\nu\sigma} - \eta_{\nu\sigma}M_{\mu\rho}\cr
&&\left[M_{\mu\nu},P_{\rho}\right]_{-}=\eta_{\nu\rho}P_{\mu} -\eta_{\mu\rho}P_{\nu}\cr
&&\left[P_{\mu},P_{\nu}\right]_{-}=M_{\mu\nu}\cr
&&\left[M_{\mu\nu},G_{\rho}\right]_{-}=\eta_{\nu\rho}G_{\mu}-\eta_{\mu\rho}G_{\nu}\cr
&&\left[P_{\mu},G_{\nu}\right]_{-}=-\eta_{\mu\nu}S_5,\quad\left[P_{\mu},G_{5}\right]_{-}=-S_{\mu},\quad\left[G_{\mu},G_{\nu}\right]_{+}=\eta_{\mu\nu}Z,\cr
&&\left[G_{\mu},G_{5}\right]_{+}=-P_{\mu},\quad\left[G_{5},G_{5}\right]_{+}=\tilde{Z}\cr
&&[G_{\mu},\tilde{Z}]_{-}=2S_{\mu},\quad\left[G_{5},Z\right]_{-}=2 S_{5},\quad
[Z,\tilde{Z}]_{-}=4Z'.
\label{VSUSYnotyet}
\end{eqnarray}
In order to perform the contraction, we rescale the $\rm OSp(3,2\,|2)$ generators with a dimensionless parameter $\lambda$ as follows:
\begin{eqnarray}
&M_{\mu\nu}\rightarrow M_{\mu\nu}, \qquad  & Z' \rightarrow Z'\cr
&P_{\mu} \rightarrow \lambda^{2} P_{\mu}, \qquad & Z \rightarrow \lambda^{2}  Z, \qquad  \tilde{Z} \rightarrow \lambda^{2} \tilde{Z} , \qquad\cr
& G_{\mu} \rightarrow \lambda G_{\mu}, \qquad & G_{5} \rightarrow \lambda G_{5} , \quad\,\, S_{\mu} \rightarrow \lambda S_{\mu}, \qquad
S_{5} \rightarrow \lambda S_{5}
\label{lambdascale}\end{eqnarray}
and consider the limit $\lambda\rightarrow\infty$. As a result, the commutation relations (\ref{VSUSYnotyet}) reduce to the VSUSY algebra.\\
In the derivation of the Casimir operators we need to eliminate the extra operators $S_{\mu}$, $S_{5}$ and $Z'$, since VSUSY is only a subalgebra of the contraction limit of $\rm OSp(3,2\,|2)$. To do that, we rescale $S_{\mu}$, $S_{5}$ and $Z'$ by a factor $\beta$ and consider the limit $\beta\rightarrow 0$.\\
$\rm OSp(3,2\,|2)$ has three independent Casimir operators \cite{Jarvis:1978bc}, $\mathcal{C}_2$, $\mathcal{C}_4$ and $\mathcal{C}_6$, constructed respectively as combinations of products of two, four and six $\rm OSp(3,2\,|2)$ generators.
For example, $\mathcal{C}_2$ reads
\begin{equation}
\mathcal{C}_2=M_{\mu\nu}M^{\mu\nu} + 2P_{\mu} P^{\mu} + 2 \left[G_{\mu},S^{\mu}\right]_{-}+2 \left[G_{5},S_{5}\right]_{-}-2Z'Z'-[Z,\tilde{Z}]_{+}\,.
\end{equation}
The direct contraction of the three independent $\rm OSp(3,2\,|2)$ Casimirs leads to combinations of $P^{2}$, $Z$ and $\tilde Z$ and explicitly one finds
\begin{equation}
\mathcal{C}_{n}\rightarrow 2\left( (P^{2})^{\frac{n}{2}} - (Z\tilde{Z})^{\frac{n}{2}}\right) \qquad n = 2,4,6.
\label{leadingCn}
\end{equation}
Since $Z$ and $\tilde Z$ are central charges, (\ref{leadingCn}) shows that $P^{2}$ is a Casimir of VSUSY.
There is no hope to derive the superspin Casimir $\hat W^{2}$ from a direct contraction procedure, since $\hat W^{\mu}$ contains $M^{\mu\nu}$, which does not scale with $\lambda$. Therefore, we construct a suitable combination of  $\mathcal{C}_2$, $\mathcal{C}_4$ and $\mathcal{C}_6$ so that the maximal order in $\lambda$ exactly cancels out. This request is uniquely satisfied by the following expression:
\begin{equation}
\mathcal{K}_{8}=-4\;\mathcal{C}_{6}\;\mathcal{C}_{2}+3\;\mathcal{C}_{4}^2+\frac{1}{4}\;\mathcal{C}_{2}^4.
\label{K8}
\end{equation}
By using the rescaled commutation relations one can prove that in $\mathcal{K}_{8}$ the maximal order $\lambda^{16}$ is vanishing and all the lower orders down to $\lambda^{13}$ are exactly zero. The first nonvanishing order is $\lambda^{12}$, which is the interesting one for the derivation of the superspin Casimir because this is the first place where $M_{\mu\nu}$ terms appear. One can check that the extra terms at order $\lambda^{12}$ generated from the higher order terms by the use of the rescaled commutation relations vanish after the limit $\lambda\to\infty$ is taken. This makes sure that they will not affect the result of the contraction. After some algebraic manipulations, one can prove the following relation
\begin{equation}
\mathcal{K}_{8}^{(12)}=48\left(P^{2}-Z\tilde{Z}\right)\left(-\frac{\tilde{Z}}{Z}\hat W^{2}+\beta^2
\{{\rm terms}~{\rm with} (S_{\mu},S_{5},Z')\}\right)+f(P^{2},Z,\tilde{Z})\,.
\label{CC812}
\end{equation}
If we take the limit $\beta\rightarrow 0$ in (\ref{CC812}), we can prove that $\hat W^{2}$ is a Casimir of VSUSY.

\section{Conclusions and outlook}
In this paper we present all independent Casimir operators of the VSUSY algebra, an odd vector extension of the four-dimensional Poincar\'e algebra. This result is the first step towards the classification of the irreducible representations of the algebra, which is currently work in progress \cite{workinprogress}.

VSUSY shares some common features with ordinary supersymmetry, for instance the fact that the anticommutator between the fermionic generators is proportional to the 4-momentum $P_{\mu}$. On the other hand, the fundamental difference is the Lorentz nature of their odd generators, spinors for ordinary supersymmetry and a vector and a scalar for VSUSY. This implies that in the case of VSUSY not only even but also odd Casimir operators may be present. We find that, in the case $Z,\tilde{Z}\neq 0$, VSUSY has four independent even Casimir operators, $\hat{W}^{2}$, $P^{2}$, $Z$ and $\tilde{Z}$. We also construct an odd operator $Q$, which behaves like a Casimir when a particular relation between the central charges and the 4-momentum is satisfied.

The Casimir operator $\hat{W}^{2}$ is the square of a Lorentz vector $\hat{W}_{\mu}$, which is the VSUSY extension of the ordinary Pauli-Lubanski vector. In the rest frame, it satisfies the $\rm SU(2)$ algebra and gives rise to the superspin $Y$, the analogue of the superspin for VSUSY. It is necessary to have both central charges different from zero to ensure that this superspin operator is an independent Casimir. On the other hand, the Casimir operator $P^{2}$ is related to another Lorentz vector, denoted by $W^{\mu}_{C}$. In the rest frame, $W^{\mu}_{C}$ also satisfies the $\rm SU(2)$ algebra and defines a different kind of spin (C-spin), fixed to the value $1/2$.

Finally, we show that VSUSY can be naturally embedded in the simple orthosymplectic superalgebra $\rm OSp(3,2\,|2)$. In order to derive the VSUSY algebra and all its independent Casimirs from $\rm OSp(3,2\,|2)$, it is necessary to introduce two different scale parameters and then perform a suitable contraction limit.

In the future, we would like to classify all irreducible representations of the VSUSY algebra. One possibility in this direction is to rewrite the odd sector of the VSUSY algebra in terms of generators of a Clifford algebra, for which all the irreducible representations have already been classified. Another possibility would be to exploit the embedding of the VSUSY algebra in $\rm OSp(3,2\,|2)$ to derive the representations.

One of our main goals would be to construct a field theoretical model displaying VSUSY invariance. A possibility would be to use a superspace approach and, in this direction, the connection between VSUSY and $\mathcal{N}=2$ topological theories could be useful, since, for the latter, a superspace setup has already been developed \cite{Kato:2005fj}. In \cite{ourpaper} we were mainly concerned with the algebraic properties of VSUSY and therefore we did not consider the issue of the violation of the spin-statistic rule mentioned in the introduction. We leave this kind of discussion for future work.

\begin{acknowledgement}
F.E. would like to thank the organizers of the 4-th EU RTN Workshop in Varna, Bulgaria, for the
opportunity to present this work. This work has been partially supported by MCYT FPA 2007-66665, CIRIT
GC 2005SGR-00564, Spanish Consolider-Ingenio 2010 Programme CPAN (CSD2007-00042),
FWO - Vlaanderen, project G.0235.05 and the Federal Office for Scientific, Technical and
Cultural Affairs through the 'Interuniversity Attraction Poles Programme – Belgian Science
Policy' P6/11-P.
\end{acknowledgement}

\end{document}